\documentclass[proceedings]{JHEP}

\usepackage{epsfig}

\def\la{\langle}
\def\ra{\rangle}

\def\l{\left}
\def\r{\right}

\def\beq{\begin{equation}}
\def\eeq{\end{equation}}
\def\bea{\begin{eqnarray}}
\def\eea{\end{eqnarray}}

\def\gev{\mbox{ GeV}}
\def\mev{\mbox{ MeV}}
\def\msbar{\overline{\mbox{MS}}}

\newcommand{\dbtwo}{\Delta B = 2}

\newcommand{\oper}{{\mathcal{O}}}

\newcommand{\ml}{{\mathcal{M}}}
\newcommand{\latt}{\mathrm{latt}}

\conference{Heavy Flavours 8, Southampton, UK, 1999}

\title{$B^{0}{-}\bar{B}^{0}$ Mixing and Decay Constants
       from Lattice QCD\thanks{CERN-TH/99-344, CPT-99/P.3909, 
Edinburgh Preprint 1999/19, LAPTH-Conf-762/99, SHEP 99/20, UK/99-16.}}

\author{L. Lellouch$^a$\thanks{Address from October, 1999: LAPTH,
        Chemin de Bellevue, B.P. 110, F-74941 Annecy-le-Vieux Cedex,
         France.  On leave from: Centre de
     Physique Th\'eorique, Case 907, CNRS Luminy, F-13288 Marseille
     Cedex 9, France.} and
      C.-J.D. Lin$^b$\thanks{Presenter at the conference.
        Address from October, 1999: Department of
      Physics $\&$ Astronomy, The University of Southampton,
      Southampton SO17 1BJ, England.}$\mbox{ }$
      (UKQCD Collaboration)\\ \ \\
     $^a$Theory Division, CERN, CH-1211 Geneva 23, Switzerland\\
     $^b$Department of Physics $\&$ Astronomy, The University
         of Edinburgh, Edinburgh EH9 3JZ, Scotland\\$\mbox{ }
         \mbox{ }\mbox{ }$and\\
         $\mbox{ }$Department of Physics $\&$ Astronomy, The University
         of Kentucky, Lexington, KY 40506-0055,\\$\mbox{ }$USA}
       \abstract{We present updates of our results for neutral
         $B$-meson mixing and leptonic
         decay constants obtained in the quenched approximation from a
         mean-field-improved Sheikholeslami-Wohlert action at two
         values of lattice spacing.  We consider quantities such as
         $B_{B_{d(s)}}$, $f_{D_{(s)}}$, $f_{B_{(s)}}$ and the full
         $\Delta B=2$ matrix-elements, as well as the corresponding
         $SU(3)$-breaking ratios.}

\begin{document}

\section{Introduction}
\label{sec:intro}

The study of $B_d-\bar B_d$ oscillations enables measurement of
the poorly known CKM matrix element $|V_{td}|$.  The frequency of
these oscillations is determined by
\begin{equation}
\Delta m_d \equiv M^{H}_{B_{d}}-M^{L}_{B_{d}} \ ,
\label{eq:dmd}
\end{equation}
where $M^{H}_{B_{d}}$ and $M^{L}_{B_{d}}$ are the heavy and light
mass eigenvalues of the mixing system.  $\Delta m_d$ is experimentally
measurable from tagged $B_{d}$ meson samples, and is also calculable
in the Standard Model.  To leading order in $1/M_W$, the 
Standard Model prediction for $\Delta m_{d}$ is \cite{Buras:1990fn,
Buchalla:1996vs}
\bea 
\Delta m_d &=& \frac{G_F^2}{8\pi^2}\, M_W^2\, |V_{td}
V_{tb}^*|^2\, \eta_B S_0(x_{t})C_B(\mu)\nonumber\\ 
\label{eq:deltamd}
&\times& \frac{1}{2 M_{B_{d}}}
|\la \bar B_d|{\mathcal{O}}^{\Delta B=2}_d(\mu)|B_d\ra|
\ ,
\eea
where $x_{t} = m^{2}_{t}/M^{2}_{W}$, $\eta_B$, $S_0(x_{t})$ and
$C_B(\mu)$ are perturbatively-calculated short-distance quantities,
$\mu$ is the renormalisation scale and ${\mathcal{O}}^{\Delta B=2}_d$
is the four-quark operator $\l[\bar b\gamma^\mu(1-\right.$
$\left.\gamma^5)d\r]$ $\l[\bar b\gamma_\mu(1-\r.$ $\l.\gamma^5)d\r]$.  Since
$|V_{tb}|$ is equal to unity to very good accuracy, a measurement of
$\Delta m_d$ enables the determination of $|V_{td}|$. The accuracy of
this determination is currently limited by the theoretical uncertainty
in the calculation of the non-perturbative strong-interaction effects
in the ma\-trix-element\hspace{0.2cm} $\la \bar B_d|$
${\mathcal{O}}^{\Delta B=2}_d$ $|B_d\ra$.  An alternative approach
\cite{BerBlumSoni}, in which many theoretical uncertainties cancel, is
to consider the ratio, $\Delta m_s/\Delta m_d$, where $\Delta m_s$ is
the mass difference in the neutral $B_s-\bar B_s$ system. In the
Standard Model, one has
\bea
\frac{\Delta m_s}{\Delta m_d}&=&
\l|\frac{V_{ts}}{V_{td}}\r|^2\frac{M_{B_s}}{M_{B_d}}\,\xi^2
=\l|\frac{V_{ts}}{V_{td}}\r|^2\frac{M_{B_d}}{M_{B_s}}\, r_{sd} 
\nonumber \\
\label{eq:deltamsovermd}
&\equiv& \l|\frac{V_{ts}}{V_{td}}\r|^2\frac{M_{B_d}}{M_{B_s}}
\l|\frac{\la \bar B_s|{\mathcal{O}}^{\Delta B=2}_s|B_s\ra}
{\la\bar B_d|{\mathcal{O}}^{\Delta B=2}_d|B_d\ra}\r|
,
\eea
where ${\mathcal{O}}^{\Delta B=2}_s$ is the same operator as
${\mathcal{O}}^{\Delta B=2}_d$ with $d$ replaced by $s$ and where we
have omitted the renormalisation-scale dependence of these operators
as it cancels in the ratio. Because the unitarity of the CKM matrix
implies $|V_{ts}|{\simeq} |V_{cb}|$ and because $|V_{cb}|$ can be
accurately obtained from semileptonic $B$ to charm decays, a
measurement of $\Delta m_s/\Delta m_d$ determines 
$|V_{td}|$. This is a challenging measurement and, at present, only a
lower bound on $\Delta m_s/\Delta m_d$ exists \cite{foa}.

\medskip

The matrix elements which appear in Eq.\ (\ref{eq:deltamsovermd}) are
traditionally parameterised by
\bea
{\mathcal{M}}_{B_q}(\mu) &=&
\la\bar B_q|{\mathcal{O}}^{\Delta B=2}_q(\mu)|B_q\ra
\nonumber \\
\label{eq:bparamdef}
&=& \frac{8}{3} M_{B_q}^2 f_{B_q}^2 B_{B_q}(\mu)
\ ,
\eea
where $q=d$ or $s$, where the $B$-parameter, $B_{B_q}$, measures
deviations from vacuum saturation, corresponding to $B_{B_q}=1$, and
$f_{B_q}$ is the leptonic decay
constant:
\beq
\la 0|\bar b\gamma_\mu\gamma_5 q|B_q(p)\ra=ip_\mu f_{B_q}
\ .\eeq
With this parameterisation, the quantity $\xi$ defined in
Eq.\ (\ref{eq:deltamsovermd}) is given by
\beq
\xi=\frac{f_{B_s}}{f_{B_d}}
\sqrt{\frac{B_{B_s}}{B_{B_d}}}
\ .
\label{eq:xidef}
\eeq
Because $M_{B_s}$ and $M_{B_d}$ are measured experimentally,
$\xi^2$
is the quantity in Eq.\ (\ref{eq:deltamsovermd}) which requires a
non-perturbative determination.

\medskip

We report on $B_{B_{d(s)}}$, $f_{B_{(s)}}$, $B_{B_{s}}/$ $B_{B_{d}}$,
$f_{B_{s}}/$ $f_{B}$, $r_{sd}$ and $\xi$.  Results for $D$-meson decay
constants $f_{D_{(s)}}$ and the $SU(3)$ breaking ratio
$f_{D_{s}}/$ $f_{D}$ are also given.  These results are updates of
those we presented in \cite{LLBBLat98,Moriond}.

\section{Main features of the lattice calculation}
\label{sec:simulation}

Numerical calculations are performed in the quenched approximation at
two values of the coupling, $\beta=6.2$ and $\beta=6.0$, corresponding
to an inverse lattice spacing $1/a \sim 2.5 \gev$ (finer) and $1/a
\sim 2.0 \gev$ (coarser), respectively.  We use a mean-field-improved
Sheikholeslami-Wohlert (SW) action \cite{BShW} to describe the quarks.
With this action, discretisation errors are formally reduced from
$\oper(a)$ to $\oper(\alpha_{s}a)$ and may be numerically smaller
because of the mean-field improvement.  This reduction of
discretisation errors is important in lattice calculations involving
heavy quarks, because these quarks have small Compton wavelengths.
For details on the parameters used in the numerical calculations,
please refer to table \ref{tab:simparam} in Appendix
\ref{sec:simdetail}.

\medskip

At each value of the lattice spacing, we have three light quarks with
masses in a range between $\sim m_{s}/2$ and $\sim m_{s}$, which
allows us to linearly extrapolate the quantities we are after to 
vanishing quark mass and interpolate them to the strange-quark mass.
To obtain results for the $b$ quark, while keeping discretisation
errors under control, we work with five heavy-quark masses straddling
the charm mass\footnote{However, only three of these are used on the
  finer lattice ($\beta=6.2$) when calculating matrix elements and
  $B$-parameters of the four-quark operators.} and extrapolate up to
the $b$ mass, guided by HQET.

\medskip

We use two methods to calculate $r_{sd}$:
\begin{itemize}
\item {\bf Direct method}: $r_{sd}$ is obtained from the direct
  calculations of $\ml_{B_{s}}$ and $\ml_{B_{d}}$.
\item {\bf Indirect method}: $r_{sd}$ is obtained by calculating
  $f_{B_{s}}/f_{B_{d}}$ and $B_{B_{s}}/B_{B_{d}}$, and then combining
  them with the experimental $M_{B_{s}}/$ $M_{B_{d}}$.
\end{itemize}
We find that both heavy-quark-mass and light-quark-mass extrapolations
are under better control for the indirect method than they are for the
direct method.

\section{Matching to the continuum and running in the $\msbar$ scheme}
\label{sec:matching}

Results of lattice-regularised calculations have to be matched to the
continuum renormalisation scheme in which Wilson coefficients are
calculated.  We perform this matching at one loop
\cite{GermanRenorm,Gabrielli,Frezzotti,Gupta} with mean-field improvement
\cite{LepageMack}.  At this order, it is consistent to use the
tree-level value for improvement coefficient $c_{\mathrm{SW}}$ (see
Appendix \ref{sec:simdetail}).  This is the procedure we use to obtain
the central values for our results.

\medskip

Moreover, because chiral symmetry is explicitly broken by Wilson
fermions, the axial vector current $A_{\mu}$ requires a
(multiplicative) renormalisation and is related to its continuum
counterpart via
\beq
 A^{\mathrm{cont}}_{\mu} = Z_{A}(\alpha_{s})
   A^{\latt}_{\mu}\mbox{,}
\label{eq:za}
\eeq
where $Z_{A}$ is finite.

\medskip

For the four-quark operators, to subtract the contributions
arising from the explicit breaking of chiral symmetry, it is sufficient to
consider the following basis of parity-conserving operators
\begin{eqnarray}
 \oper_{1,2} &=& \gamma_{\mu}\times\gamma_{\mu} \pm
     \gamma_{\mu}\gamma_{5}\times\gamma_{\mu}\gamma_{5}\mbox{,} \nonumber \\
 \oper_{3,4} &=& I\times I \pm \gamma_{5}\times\gamma_{5}\mbox{,} \\
\label{eq:4q_renorm}
 \oper_{5} &=& \sigma_{\mu\nu}\times\sigma_{\mu\nu}\mbox{,} \nonumber
\end{eqnarray}
where we only show their Dirac structure for simplicity.  $\oper_1$
is the parity-even part of $\oper^{\dbtwo}_q$.  This operator, in the
$\msbar$ scheme at the renormalisation scale $\mu$, is related to
the above lattice operators by
\beq \oper^{\tiny{\msbar}}_{1}(\mu) = Z_{11}(\alpha_{s},a\mu)
\hat{\oper}^{\mathrm{latt}}_{1}(a)\ ,
\eeq
where
\beq
   \hat{\oper}^{\mathrm{latt}}_{1}(a) = \oper^{\latt}_{1}(a) +
   \sum^{5}_{i=2} Z_{1i}(\alpha_{s})
                    \oper^{\latt}_{i}(a)\ .
\eeq
$Z_{11}$ depends logarithmically on $a\mu$.  The $Z_{1i}$, $i\ne 1$,
account for the operator mixing due to the explicit chiral symmetry,
and do not depend on $a\mu$.

\medskip

The scheme of $\alpha_s$ is not fixed at one loop. We choose
$\alpha_s=\alpha_{\tiny{\msbar}}$ obtained from the procedure
described in \cite{LepageMack}, with $n_f=0$ (quenched approximation),
which was shown to lead to particularly convergent perturbative
expansions \cite{LepageMack}.  Central values are obtained by
identifying the scale of the coupling with the matching scale $\mu$
and matching at $\mu=2/a$--a typical lattice ultraviolet scale.
Running in the $\msbar$ scheme is performed at two loops with the same
coupling constant as for the matching, and $n_f=0$.

\section{Scaling with heavy-quark mass}
\label{sec:hq}
\FIGURE{
\epsfig{file=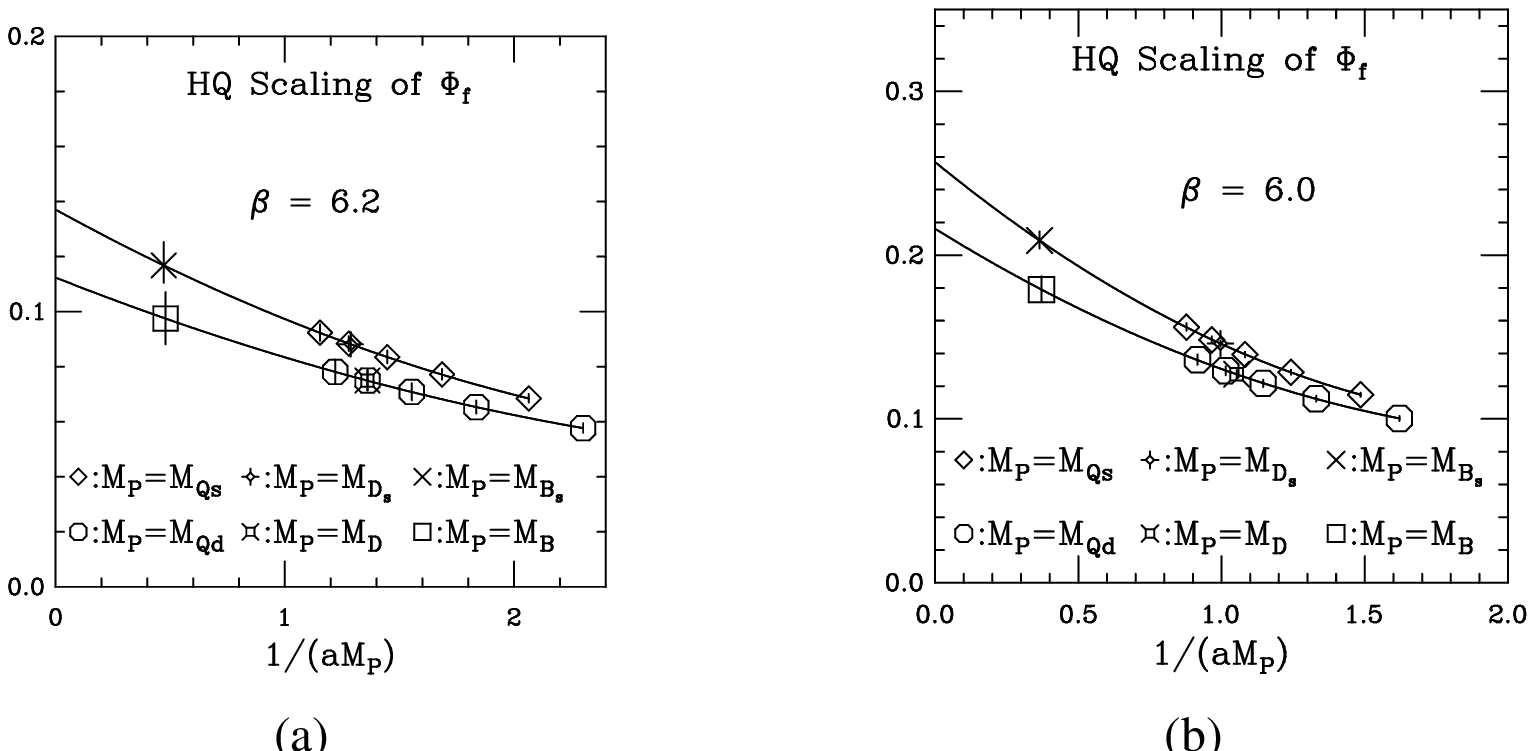,height=6.6cm,width=13.2cm}
\caption{\sl Scaling of
  $\Phi_{f}$ with heavy-quark mass on (a): the finer lattice and 
 (b): the coarser lattice.  The points labelled $M_{Qs}$ and
  $M_{Qd}$ correspond to the heavy quarks, $Q$, used in our
  simulation. The curves are fits to the RHS of Eq.\ 
  (\ref{eq:hqscal}). The other points are the result of
  interpolation to $Q=c$ or extrapolation to $Q=b$.}
\label{fig:fhq}}
\FIGURE{
\epsfig{file=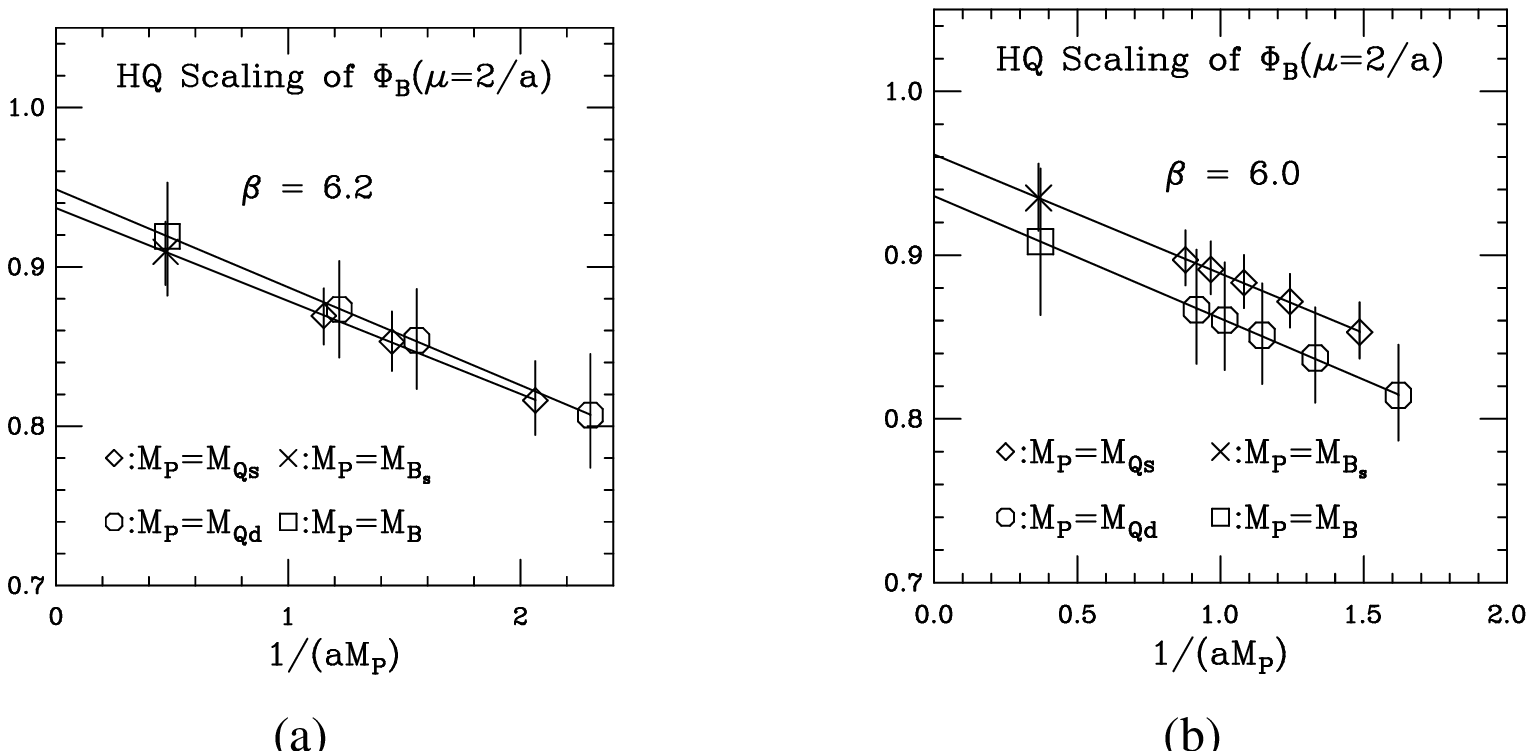,height=6.6cm,width=13.2cm}
\caption{\sl Scaling of
  $\Phi_{B}$ with heavy-quark mass on (a): the finer lattice and 
  (b): the coarser lattice.  The points labelled $M_{Qs}$ 
  and $M_{Qd}$ correspond to the heavy quarks, $Q$, used in our
  simulation. The curves are fits to the RHS of Eq.\ 
  (\ref{eq:hqscal}) without the quadratic term in $1/M_P$.  The
  other points are the result of extrapolation to $Q=b$.}
\label{fig:Bhq}}
\FIGURE{
\epsfig{file=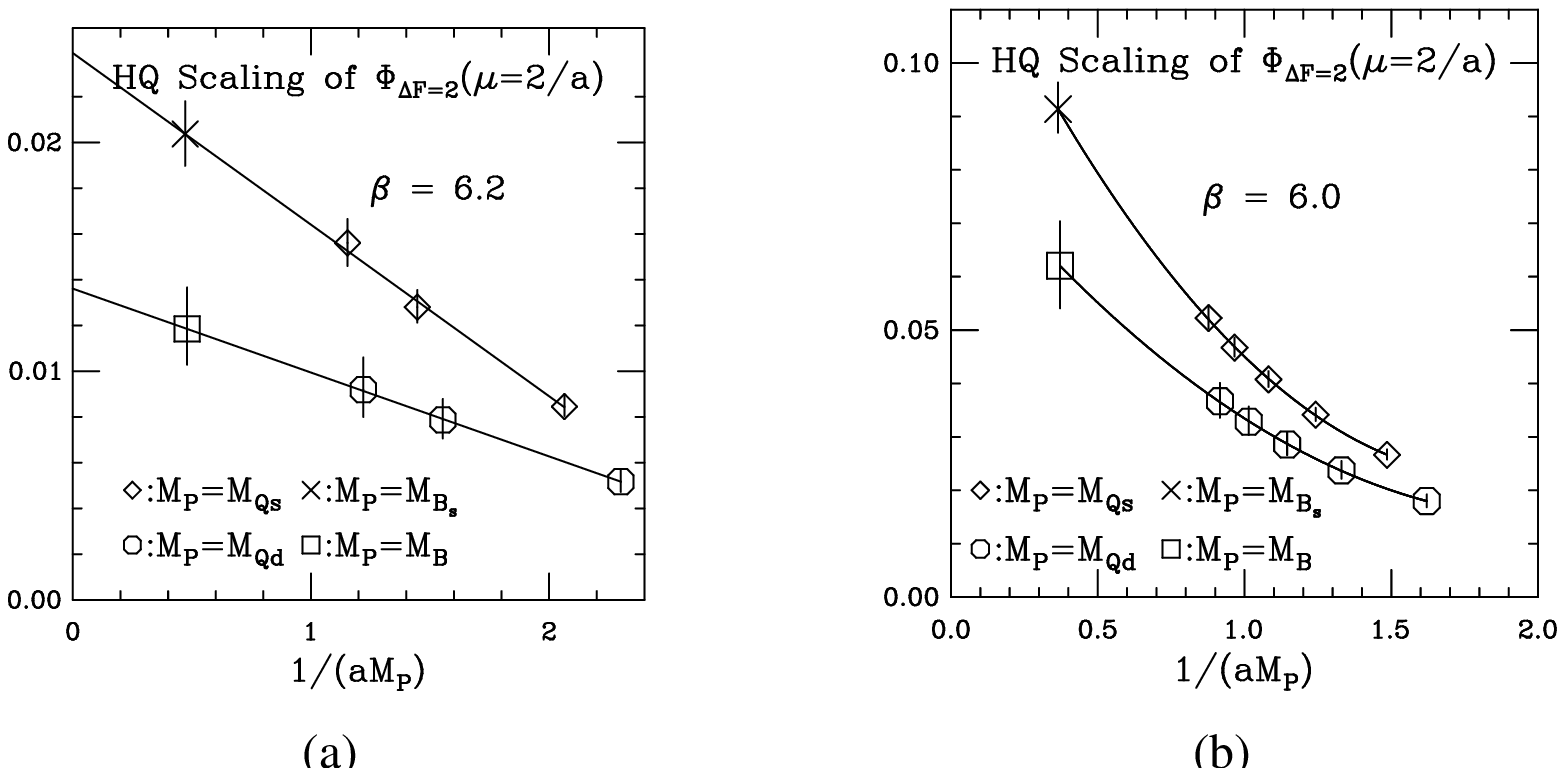,height=6.6cm,width=13.2cm}
\caption{\sl Scaling of
  $\Phi_{\Delta F = 2}$ with heavy-quark mass on (a): the finer lattice and
  (b): the coarser lattice.  The points labelled
  $M_{Qs}$ and $M_{Qd}$ correspond to the heavy quarks, $Q$, used in
  our simulation. The curves are fits to the RHS of Eq.\ 
  (\ref{eq:hqscal}) (without the quadratic term in $1/M_P$ for the 
  finer lattice).  The
  other points are the result of extrapolation to $Q=b$.}
\label{fig:mlhq}}

To study the behaviour of the various quantities with heavy-quark mass,
we define
\beq
\label{eq:phifdef}
 \Phi_{f}(M_{P}) \equiv
\frac{af_{P}}{Z_A}\sqrt{aM_{P}}
\l(\frac{\alpha_s(M_P)}{\alpha_s(M_B)}\r)^{2/\beta_0}
\eeq
\beq
\label{eq:phiBdef}
 \Phi_{B}(\mu,M_{P}) \equiv
B_{P}(\mu)\l(\frac{\alpha_s(M_P)}{\alpha_s(M_B)}\r)^{2/\beta_0}
\eeq
\beq
\label{eq:phidf2def}
\Phi_{\Delta F=2}(\mu,M_P)  \equiv
\frac{a^4{\mathcal{M}}_{P}(\mu)}{aM_{P}}
\l(\frac{\alpha_s(M_P)}{\alpha_s(M_B)}\r)^{6/\beta_0}
\eeq
where $M_{P}$ is the heavy-meson mass and $\ml_{P}$ is the $\Delta F =
2$ matrix element calculated at that mass. $\beta_{0}$ is the one-loop
coefficient of the QCD $\beta$-function, with $n_{f}=0$.  In
$\Phi_{f}$, $\Phi_{B}$ and $\Phi_{\Delta F = 2}$, we have cancelled
the logarithmic dependence of $f_P$, $B_P$ and $\ml_{P}$ on $M_{P}$ at
leading-log order \cite{PolitzerWise}.

\medskip

For $X(M_P)=\Phi_{f}$,
$\Phi_{B}$, $\Phi_{\Delta F=2}$ and the
$SU(3)$-breaking ratios,  we use the HQET-inspired relation,
\beq
X(M_P) =
a_X+b_X\l(\frac{1}{aM_P}\r)+c_X\l(\frac{1}{aM_P}\r)^2+\cdots,
\label{eq:hqscal}
\eeq
to investigate the heavy-quark-mass scaling behaviour of these quantities,
as shown in figures \ref{fig:fhq}, \ref{fig:Bhq} and \ref{fig:mlhq}.

\bigskip

\section{Systematic uncertainties}
\label{sec:results}

\TABLE[t]{
\begin{tabular}{|c|c|c|}
\hline
lattice & coarser & finer\\
\hline
$\beta$ & 6.0 & 6.2\\
\hline
$f_{D_{s}} (\mathrm{MeV})$ &
$251(3)^{+16+11+2+17}_{-\phantom{1}0-\phantom{1}4-0-\phantom{1}8}$ 
& $224(7)^{+9+8+0+16}_{-0-3-0-\phantom{1}8}$
\\ \hline
$f_{D} (\mathrm{MeV})$ &
$224(4)^{+12+10+1+20}_{-\phantom{1}0-\phantom{1}4-0-10}$ 
& $195(10)^{+7+7+0+19}_{-0-3-0-10}$
\\ \hline
$f_{B_{s}} (\mathrm{MeV})$ &
$232(6)^{+27+10+\phantom{1}0+22}_{-\phantom{1}0-\phantom{1}4-16-11}$ & 
$192(14)^{+14+7+\phantom{1}0+19}_{-\phantom{1}0-3-10-\phantom{1}9}$
\\ \hline
$f_{B} (\mathrm{MeV})$ &
$201(9)^{+20+9+\phantom{1}0+25}_{-\phantom{1}0-4-13-13}$ & 
$161(16)^{+11+6+0+21}_{-\phantom{1}0-2-8-10}$
\\ \hline
$f_{D_{s}}/f_{D}$ &
$1.12(1)^{+1+\phantom{1}+\phantom{1}+2}_{-0-\phantom{1}-\phantom{1}-2}$ 
& $1.15(4)^{+1+\phantom{1}+\phantom{1}+2}_{-0-\phantom{1}-\phantom{1}-3}$
\\ \hline
$f_{B_{s}}/f_{B}$ &
$1.14(2)^{+1+\phantom{1}+\phantom{1}+2}_{-0-\phantom{1}-\phantom{1}-3}$ & 
$1.16(6)^{+1+\phantom{1}+\phantom{1}+2}_{-0-\phantom{1}-\phantom{1}-3}$
\\ \hline
$B_{B_{s}}(5\gev)$ &
$0.92(2)^{+\phantom{1}+4+\phantom{1}+
0}_{-\phantom{1}-0-\phantom{1}-0}$ 
& $0.91(2)^{+\phantom{1}+3+\phantom{1}+
0}_{-\phantom{1}-0-\phantom{1}-0}$
\\ \hline
$B_{B}$(5\gev) &
$0.90(4)^{+\phantom{1}+4+\phantom{1}+
0}_{-\phantom{1}-0-\phantom{1}-0}$ 
& $0.92(4)^{+\phantom{1}+3+\phantom{1}+
0}_{-\phantom{1}-0-\phantom{1}-0}$
\\ \hline
$B_{B_{s}}/B_{B_{d}}$
& $1.02(3)^{+\phantom{1}+0+\phantom{1}+
0}_{-\phantom{1}-0-\phantom{1}-0}$ & 
$0.98(3)^{+\phantom{1}+0+\phantom{1}+
0}_{-\phantom{1}-0-\phantom{1}-0}$
\\ \hline
$r^{\mathrm{indirect}}_{sd}$ &
$1.38(7)^{+2+\phantom{1}+\phantom{1}+
4}_{-0-\phantom{1}-\phantom{1}-6}$ 
& $1.37(14)^{+1+\phantom{1}+\phantom{1}+
4}_{-0-\phantom{1}-\phantom{1}-6}$
\\ \hline
$r^{\mathrm{direct}}_{sd}$ &
$1.52(18)^{
+\phantom{1}+\phantom{1}+\phantom{1}+
6}_{-\phantom{1}-\phantom{1}-\phantom{1}-9}$ 
& $1.71(28)^{+\phantom{1}+\phantom{1}+\phantom{1}+
\phantom{1}8}_{-\phantom{1}-\phantom{1}-\phantom{1}-11}$
\\ \hline
$\xi$ & $1.15(3)^{+1+\phantom{1}+\phantom{1}
+2}_{-0-\phantom{1}-\phantom{1}-3}$ & 
$1.15(6)^{+1+\phantom{1}+\phantom{1}+
2}_{-0-\phantom{1}-\phantom{1}-3}$\\ \hline
\end{tabular}
\caption{\sl Results at the two values of
lattice spacing. $r^{\mathrm{indirect}}_{sd}=
(\frac{M_{B_{s}}}{M_{B}}\frac{f_{B_{s}}}{f_{B}})^{2}
\frac{B_{B_{s}}}{B_{B}}$ and $r^{\mathrm{direct}}_{sd}=
({\mathcal{M}}_{B_{s}}/{\mathcal{M}}_{B_{d}})$.  The first error bar
on each quantity is statistical while the others are systematic, as
described in Section \protect\ref{sec:results}. Blank error bars are put in
to help keep track of which systematic effect each error corresponds to.}
\label{tab:results}}
Our main results at the two values of lattice spacing are summarised
in table \ref{tab:results}.  In this table, the first error bar for
each quantity is statistical.  The other errors are systematic and we
discuss them now.

\subsection{Discretisation errors}

In table \ref{tab:results}, results for the decay constants display
significant variation with lattice spacing\footnote{This poor scaling
  is not fully understood. It could be improved by using $f_\pi$
  instead of $m_\rho$ to set the scale (see table \ref{tab:simparam}).
  For instance, $f_B$ at $\beta=6.0$ would be $\sim 191\mev$ instead
  of $201\mev$ while its value at $\beta=6.2$ would be $160\mev$.
  However, $m_\rho$ is a valid means of setting the scale in quenched
  calculations and, as discussed below, we include a systematic
  associated with the uncertainty in the lattice spacing. Furthermore,
  this poor scaling is not present in the $B$-parameters and
  $SU(3)$-breaking ratios which are the main thrust of the present
  work.}. This suggests that discretisation errors for these
quantities may be important. To quantify these errors we estimate
residual, ${\mathcal{O}}(am_Q\alpha_{s})$ discretisation effects,
associated with the mass $m_Q$ of the heavy quark, as described in
Appendix \ref{sec:improveA}. This is the second error bar on the
decays constants and their $SU(3)$-breaking ratios, $f_{D_{s}}/f_{D}$
and $f_{B_{s}}/f_{B}$.

\medskip

Such an estimate could, in principle, be carried out for the
$B$-parameters and the $SU(3)$-breaking ratios $r_{sd}$ and $\xi$.
However, many of these discretisation effects cancel trivially in the
ratios of matrix elements defining these quantities.  Furthermore, a
full quantification of $\mathcal{O}(a$ $m_Q\alpha_{s})$ effects for
$\ml_{B_{q}}$ and their $B$-parameters would require one to consider
the mixing of the four-quark operators in Eq.\ (\ref{eq:4q_renorm})
with operators of dimension seven, which is beyond the scope of the
present work. Finally, in table \ref{tab:results}, results for
$B$-parameters and their $SU(3)$-breaking ratios exhibit very little
lattice-spacing dependence, supporting the idea that discretisation
errors for these quantities are small. Thus, we assume that the
statistical error for these quantities encompasses possible residual
discretisation errors. For $r^{\mathrm{indirect}}_{sd}$ and $\xi$,
however, which are obtained using $f_{B_{s}}/f_{B}$, we take into
account the discretisation error on this quantity.

\subsection{Matching uncertainties}

To estimate the systematic errors arising from the perturbative
matching in the $B$-parameters, we match at different $\mu$ in the
range between $1/a$ and $\pi/a$ \footnote{We always identify the scale
of $\alpha_{\tiny{\msbar}}$ with the matching scale.}, then run the
resultant $B$-parameters to $2/a$ to compare them with the ones
matched ``directly'' at $2/a$. The range $[1/a, \pi/a]$ covers
typical lattice ultraviolet scales and is vindicated by our study of
$B_{K}$ \cite{LandLBK}, where we find that continuum
chiral behaviour is restored for these scales. We also consider the
variation coming from computing $Z_{11}(\mu=2/a)$ and $Z_{1i}$ with
the constant $c_{\mathrm{SW}}$ set to its mean-field-improved value
instead of 1. All of these variations, which affect $B_B$
and $B_{B_{s}}$ significantly, but not $B_{B_{s}}/$ $B_B$, are
reflected in the second error bars on these $B$-parameters.

\medskip

Decay constants are independent of renormalisation scale. However, the
above procedure results in a $\sim 4\%$ change in $Z_{A}$ through the
$\mu$-dependence of $\alpha_{\tiny{\msbar}}(\mu)$ and the change in
the value of $c_{\mathrm{SW}}$. This is reflected in the decay
constants' third error bar but does not affect the corresponding
$SU(3)$-breaking ratios.

\subsection{Heavy-quark-mass extrapolations}

As shown in figure \ref{fig:fhq}, the decay constants have a
pronounced extrapolation in heavy-quark mass, and the term quadratic
in $1/M_P$ on the RHS of Eq.\ (\ref{eq:hqscal}) contributes
significantly.  To quantify the systematic error associated with this
extra\-polation--the fourth error bar on the decay con\-stants--we perform
a fit of the heaviest three points in figure \ref{fig:fhq} to the RHS
of Eq.\ (\ref{eq:hqscal}), without the quadratic term.

\medskip

Figure \ref{fig:Bhq} indicates that the linear heavy-quark-mass
extrapolation of the $B$-parameters works well and is mild: the
associated uncertainty should be covered by the statistical error.

\medskip

The $\Delta F = 2$ matrix elements have a very pronounced dependence
on heavy-quark mass, as seen in figure \ref{fig:mlhq}. Since we are
not reporting results for the individual $\ml_{B_{d(s)}}$, we do not
quantify the systematic errors associated with their determination.
However, this strong mass-de\-pen\-dence is one of the elements which make
a reliable determination of $r_{sd}$, from the ratio of individually
calculated $\ml_{B_{d(s)}}$, difficult \cite{LandLPaper}.

\subsection{Uncertainties in the determination of the lattice spacing}

In quenched calculations, the value of the lattice
spacing varies significantly with the quantity used to set the scale.
This variation may be due to quenching effects, as well as any other
systematic uncertainty which may affect the quantity used to set the
scale. In this work, we determine the lattice spacing from the
$\rho$-meson mass\footnote{The scale determined from $f_{\pi}$ gives
compatible results at $\beta=6.2$, as shown in table 
\ref{tab:simparam}.}.  We then
vary the inverse lattice spacing, $1/a$, increasing it by $10\%$
and decreasing it by $5\%$.  This range covers the typical variations
observed in the determination of the scale from gluonic or light-hadron
spectral quantities, for the action and parameters we use
\cite{Bowler:1999ae}.

\medskip

Uncertainties in the lattice spacing will obviously affect the
determination of all the decay constants, as they are dimensionful.
They will also slightly change the curves in the heavy-quark-mass
extrapolations (figures \ref{fig:fhq}, \ref{fig:Bhq} and
\ref{fig:mlhq}).  Furthermore, they induce a variation in the
stran\-ge-quark mass, which we obtain from the mass of the kaon, and
therefore affect all quantities which depend on this mass.

\medskip

In practice, we find that the variation of the lattice spacing
discussed above does not induce a significant change in the
$B$-parameters.  However, it does affect all the decay constants and
$SU(3)$-breaking ratios.  This is reflected in the last error bar on
these quantities.

\subsection{Quenching errors}

Quenching errors have been studied using quenched Chiral Perturbation
Theory (q$\chi$PT) and have been found to be small for the
$B$-parameters \cite{QChPT}. Moreover, numerical simulation with two
light flavours of dynamical quarks indicate that those in
$f_{B_{s}}/f_{B}$ are also small \cite{MILCfB,Shoji}. Thus, they ought
to be small for $r_{sd}$ and $\xi$.  Contrary to this, quenching
errors may be significant for the decay constants themselves, as
indicated by both q$\chi$PT \cite{QChPT} and numerical simulation
\cite{MILCfB,Shoji}.

\medskip

As we mentioned in the previous section, the uncertainty in the
lattice scale may be, in part, a quenching effect. Thus, to the extent
that it is, we have already accounted for some quenching errors.  A
more thorough estimate of these effects, however, would require a
dedicated unquenched simulation which is beyond the scope of this work.
Therefore, we do not attempt to quantify quenching errors any further.

\section{Final results}

Since two lattice spacings are not sufficient for an extrapolation to
the continuum limit ($a=0$), we quote the results obtained on the
finer lattice ($\beta=6.2$) as our best estimates. And because it
appears to be more reliable, we quote $r^{\mathrm{indirect}}_{sd}$ for
$r_{sd}$. 

\medskip

Our main preliminary results are thus
\bea
 &&\xi = 1.15(6)^{+2}_{-3} \ , \nonumber   \\
 &&r_{sd} = 1.37(14)^{+4}_{-6}\ , \nonumber  \\
 &&\frac{B_{B_{s}}}{B_{B}} = 0.98(3)\ , \nonumber  \\
 &&\frac{f_{B_{s}}}{f_{B}} = 1.16(6)^{+2}_{-3}\ , \nonumber  \\
 &&B_{B}(5\gev) = 0.92(4)^{+3}_{-0}\ , \nonumber  \\
 &&B_{B_{s}}(5\gev) = 0.91(2)^{+3}_{-0}\ , \nonumber \\
 &&f_{B} = 161(16)^{+24}_{-13}\mbox{ }\mev\ , \nonumber  \\
 &&f_{B_{s}} = 192(14)^{+24}_{-13}\mbox\mev\ , \nonumber  \\
 &&f_{D} = 195(10)^{+22}_{-10}\mbox{ }\mev\ , \nonumber  \\
 &&f_{D_{s}} = 224(7)^{+21}_{-9}\mbox{ }\mev\ , \nonumber  \\
 &&\frac{f_{D_{s}}}{f_{D}} = 1.15(4)^{+2}_{-3}\ , \nonumber  
\eea
where the first error bar is statistical and the second is systematic,
the result of adding our long list of systematic errors in quadrature.
By choosing the results obtained on the finer lattice, we are also
being conservative in our estimate of statistical errors as they are
larger than on the coarser lattice for which we have higher statistics.

\medskip

Note that our results, obtained with large statistics and a highly
improved action are compatible with recent world averages
\cite{Moriond,FlynnSachrajda,Terry,Shoji}.

\acknowledgments

This work is supported in part by EPSRC and PPARC under grants
GR/K41663 and GR/L29\-927.  We warmly thank Damir Becirevic, Gerhard
Buchalla, Jonathan Flynn, Vittorio Lubicz and Guido Martinelli for
informative discussions and correspondance. We are grateful to the UKQCD
Collaboration, especially Peter Boyle, Pablo Martinez, Chris Maynard
and David Richards, for their help.  L.L. wishes to thank the
Department of Physics $\&$ Astronomy of the University of Edinburgh
and C.-J.D.L. wishes to thank CPT Marseille and the CERN Theory
Division, for hospitality.  Financial support from the DOE EPSCoR
grant DE-FG05-84ER40154 (PI Liu, CO-PI Draper) is also acknowledged by
C.-J.D.L..

\appendix

\section{Simulation Details}
\label{sec:simdetail}
The quarks in the simulation are described by the SW action
\beq
S_{\mathrm{F}}^{\mathrm{SW}} = S_{\mathrm{F}}^{\mathrm{W}} -
ig_0\,c_{\mathrm{SW}}\frac{\kappa_q}{2}\sum_{x,\mu,\nu}\,
            \bar q\,F_{\mu\nu}\sigma_{\mu\nu}\,q(x)\,
\ ,
\label{eq:sw}
\eeq
where $S_{\mathrm{F}}^{\mathrm{W}}$ is the standard Wilson action,
$g_0$ the bare gauge coupling, $F_{\mu\nu}$ a lattice realisation of
the Yang-Mills field strength tensor, $\kappa_q$ the appropriate quark
hopping parameter and $c_{\mathrm{SW}}$, the so-called clover
coefficient.  $c_{\mathrm{SW}}=1$ corresponds to tree-level
improvement. With this value of $c_{\mathrm{SW}}$, leading
discretisation errors are $\oper(\alpha_{s}a)$ instead of $\oper(a)$
as they are with the standard Wilson action, corresponding to
$c_{\mathrm{SW}}=0$. We actually use a mean-field-improved SW action
with values of $c_{\mathrm{SW}}$ given in table \ref{tab:simparam},
where the parameters used in our simulations are
summarised.
\TABLE[t]{
\begin{tabular}{|c|c|c|}
\hline
lattice & coarser & finer\\

\hline
$\beta$ & 6.0 & 6.2\\
\hline
lattice size & $16^{3}\times 48$ & $24^{3}\times 48$\\
\hline
$c_{\mathrm{SW}}$ & 1.47852 & 1.44239\\
\hline
$\#$ of cfs. & 498 & 188\\
\hline
$a^{-1}(M_{\rho})$ (GeV) & 1.96(5) & 2.54(8)\\
\hline
$a^{-1}(f_{\pi})$ (GeV) & 1.87(4) & 2.52(8)\\
\hline
\end{tabular}
\caption{\sl Simulation parameters. $a^{-1}(M_{\rho})$
  and $a^{-1}(f_{\pi})$ are the values of the inverse lattice spacing
  determined from calculations of the $\rho$-meson mass and the pion
  decay constant, respectively. The latter is given for $\mu=2/a$.}
\label{tab:simparam}}

\section{$\oper(a)$-improvement of the axial current}
\label{sec:improveA}

The leading discretisation errors with the mean-field-improved SW
action are {\it formally} of $\oper(\alpha_{s}a)$, as they are for the
tree-level improved SW action. To estimate these errors, we consider
the following variation in our procedure.

\medskip

$\oper(\alpha_{s}a)$-improvement of the axial current requires
one to include the effect of the
$a\partial_{\mu}P$ ($P$ the pseudoscalar
density) counterterm through the replacement
\beq
 A_{\mu}\to A_{\mu} + c_{A}a\partial_{\mu}P
\ ,\eeq
as well as to rescale the quark fields as
\beq
q \to (1 + \frac{b_{A}}{2}
 am_q)q
\label{eq:qnorm}\ ,\eeq
with both $c_A$ and $b_A$ evaluated at one loop
\cite{alpha,oneloopimp}.  Thus, from a comparison of results obtained
with $c_A$ and $b_A$ set to their tree-level values ($c_A=0$ and
$b_A=1$) to those obtained with $c_A$ and $b_A$ evaluated at one loop,
we can have an estimate of the effect of $\oper(\alpha_{s}a)$
discretisation errors. We do not use the one-loop results as central
values for the decay constants, for consistency with our determination of
the $B$-parameters. Indeed, $\oper(\alpha_{s}a)$-improvement of the
four-quark operators would require one to consider the mixing of these
operators with operators of dimension seven, which is beyond the scope
of the present work.

\medskip

To correct for some higher-order discretisation effects, we actually
use KLM normalisation \cite{KLM} for the quark fields. Thus, our central
values are obtained with the normalisation
\beq
q \to \sqrt{1 +  am_q}q
\eeq
and the one-loop variation with
\beq
q \to \frac{\sqrt{1 +  am_q}}{1 +  am_q/2}(1 + 
\frac{b_{A}^{\mathrm{1-loop}}}{2}
 am_q)q
\ .\eeq
We also check that results obtained with tree-level normalisation
($b_A=1$ in Eq.\ (\ref{eq:qnorm})) \cite{Impr} lie within the
discretisation error we quote.

\end{document}